\documentclass[12pt]{article}
\usepackage{
amsfonts,comment,amsmath,amssymb, 
geometry,esint}
\catcode`\@=11

\@addtoreset{equation}{section}

\usepackage{color}

\usepackage{mathtools}

\def\SO{SO(3)}
\def\Id{\text{Id}_3}
\def\mR{{R}}

\allowdisplaybreaks

\def\R{\mathbb{R}}

\def\ms{~\!}
\def\deg{\rm deg}
\def\Sf{\mathbb{S}}

\def\eps{\varepsilon}
\def\beps{\overline\varepsilon}

\def\irn{\int\limits_{\S^1}\!\!}

\def\fintm{\fint\limits_{\S^1}}

\def\proof{\noindent{\textbf{Proof. }}}
\def\QED{\hfill {$\square$}\goodbreak \medskip}

\newtheorem{Theorem}{Theorem}[section]
\newtheorem{Lemma}[Theorem]{Lemma}

\newtheorem{Remark}[Theorem]{Remark}
\newtheorem{Definition}[Theorem]{Definition}

\begin{document}

\title{
{Many  closed $K$-magnetic geodesics on $\Sf^2$ }
\author{Roberta Musina\footnote{Dipartimento di Scienze Matematiche, Informatiche e Fisiche, Universit\`a di Udine,
via delle Scienze, 206 -- 33100 Udine, Italy. Email: {roberta.musina@uniud.it}. 
{Partially supported by PRID project VAPROGE.}} \and 
Fabio Zuddas\footnote{Dipartimento di Matematica e Informatica, Universit\`a di Cagliari, via Ospedale, 72 -- 09134 Cagliari, Italy. Email: {fabio.zuddas@unica.it}. 
{Supported by Prin 2015 – Real and Complex Manifolds; Geometry, Topology and Harmonic Analysis – Italy, by STAGE - Funded by Fondazione di Sardegna and by KASBA- Funded by Regione Autonoma della Sardegna.}}}}
\date{}

\maketitle

\begin{abstract}{


In this paper we adopt an alternative, analytical approach to Arnol'd problem \cite{A1} about the existence of 
closed and embedded $K$-magnetic geodesics in the round $2$-sphere $\Sf^2$, where $K: \Sf^2 \rightarrow \R$ is a smooth scalar function. 
In particular, we use Lyapunov-Schmidt finite-dimensional reduction coupled with a 
local variational formulation in order to get some existence and multiplicity results bypassing the use of symplectic geometric tools such as the celebrated Viterbo's theorem  \cite{V} 
and Bottkoll results \cite{Bo}. }
\end{abstract}

\section{Introduction}

We deal with the motion $\gamma=\gamma(t)$ of a particle of unit mass
and charge in $\R^3$, that experiences the Lorentz force 
${\bf F}$ produced by a magnetostatic field $\bf B$. If the particle is constrained to 
the standard round sphere $\Sf^2\subset \R^3$, the motion law reads
\begin{equation}
\label{eq:Arnold}
\gamma''+|\gamma'|^2\gamma=K(\gamma)\ms \gamma\wedge\gamma'\ms,
\end{equation}
where 
$$
K(p):=-{\bf B}(p)\cdot p~,\quad p\in\Sf^2~\!.
$$
A trajectory $\gamma(t)$ satisfying  (\ref{eq:Arnold}) is called {\em $K$-magnetic geodesic}. 

Let us recall the elementary derivation of (\ref{eq:Arnold}).  We have 
${\bf F}(\gamma)=\gamma'\wedge {\bf B}(\gamma)$; due to the constraint $|\gamma|\equiv 1$, the vectors $\gamma$ and $\gamma'$ are orthogonal along the motion.
It follows that the projection of ${\bf F}$ on $T_\gamma\Sf^2=\langle\gamma\rangle^\perp$ is 
proportional to $\gamma\wedge \gamma'$, and in fact ${\bf F}^{T\!}(\gamma)=-({\bf B}(\gamma)\cdot\gamma)\ms\gamma\wedge\gamma'=K(\gamma)\ms\gamma\wedge\gamma'$. Finally, by differentiating the identity
$\gamma\cdot\gamma'\equiv 0$, we see that the tangent component of the acceleration vector is 
$\gamma''-(\gamma''\cdot\gamma)\gamma=\gamma''+|\gamma'|^2\gamma$, and thus
Newton's law gives (\ref{eq:Arnold}).
Notice that $\gamma''-(\gamma''\cdot\gamma)\gamma=\nabla^{\Sf^2}_{\!\gamma'}\gamma'$, where $\nabla^{\Sf^2}$ is the Levi-Civita
connection of $\Sf^2$.

Two remarkable facts immediately follow  from (\ref{eq:Arnold}). First, we have
$2\gamma''\cdot\gamma'=(|\gamma'|^2)'=0$. Thus the particle moves with constant scalar speed, say
$$
|\gamma'|\equiv c\ms,
$$
for some  $c>0$. In particular, $\gamma$ is a regular curve. Secondly, we learn from 
differential geometry that 
$\gamma$ has geodesic curvature
$$
\kappa(\gamma)=\frac{\gamma''\cdot \gamma\wedge\gamma'}{|\gamma'|^3}=\frac{K(\gamma)}{c}\ms.
$$

Next, let $c>0$ and $K:\Sf^2\to \R$ be given. In \cite{A1}, see also \cite[Problems 1988/30, 1994/14, 1996/18]{A2},  Arnol'd 
proposed the following
question (actually in a more general setting, where $\Sf^2$ is replaced by an oriented Riemannian surface $(\Sigma,g)$):
\begin{equation*}\tag{$\mathcal P_{K,c}$}
\label{P_Arnold}
\begin{aligned}
&\text{\em Find  closed and embedded $K$-magnetic geodesics $\gamma\subset\Sf^2$ with  $|\gamma'|\equiv c$.}
\end{aligned}
\end{equation*}
Problem (\ref{P_Arnold}), together with its generalizations, attracted the attention of many authors and has been studied
via different mathematical tools, such as
symplectic geometric
\cite{A1, G1, G2, GG2, Sl} and variational arguments for multivalued functionals \cite{BT, NT, T1, T2}. 


The relation between Problem (\ref{P_Arnold}) and symplectic geometry can be explained as follows. Let us consider on $\Sf^2$ the (restriction of the) two-form $\beta := i_{\bf B}(dx \wedge dy \wedge dz)$ and let us define on the cotangent bundle $T^* \Sf^2$ endowed with coordinates $(q, p)$ the symplectic form
$$\Omega = c~ dq \wedge dp - \pi^* \beta$$
where $dq \wedge dp = \sum_{i=1}^2 dq_i \wedge dp_i$ denotes the {\it standard symplectic form} on $T^* \Sf^2$ and $\pi: T^* \Sf^2 \rightarrow \Sf^2$ is the canonical projection.

It is not hard to show, via a straight calculation, that  $K$-magnetic geodesics on $\Sf^2$ having constant speed $c$ are exactly the projections $\pi(\gamma)$ of the integral curves of the vector field on $T^* \Sf^2$ defined by

\begin{equation}\label{dH}
d H = i_X \Omega~\!,
\end{equation}
where $H = \frac{1}{2} |p|^2$. 
In the language of symplectic geometry, $X$ is the {\it Hamiltonian vector field} given by the Hamiltonian function $H$. Notice also that since $\gamma'$ as observed above has constant speed, then $H(\gamma)$ is constant and then by (\ref{dH}) we have $i_{\gamma'} \Omega = 0$, which by definition means that $\gamma$ is a {\it characteristic} of $\Omega$.

Now, for any smooth $K$ and every $c > 0$ large enough the existence of a solution to (\ref{P_Arnold}) can be deduced via this symplectic geometric approach  by applying the celebrated Viterbo result \cite{V} on the existence of closed characteristics on compact hypersurfaces  of contact type. It is worth to notice that this result can be generalized to any closed oriented surface $\Sigma$, yielding the existence of a solution for high energies $c$ in every free homotopy class that can be represented by a non-degenerate geodesic \cite[Theorem 2.1 (ii)]{G2}.

For the case of low energy levels we cite \cite[Theorem 2.1 (i)]{G2} and \cite{Sl}, where the author
proves the existence of contractible periodic solutions for almost all sufficiently small energy
levels and for arbitrary smooth magnetic fields.

The existence of at least two distinct\footnote{\footnotesize{We agree that the curves $\gamma_1(t), \gamma_2(t)$ are distinct if 
$\gamma_1\neq \gamma_2\circ g$, for any diffeomorphism $g$.}} solutions to (\ref{P_Arnold}) in the case of the round two-sphere follows, always for $c > 0$ large enough, from a general result of Bottkoll
\cite{Bo} (see also \cite{AbbBe}) about the number of periodic orbits of the flow of a Hamiltonian vector field which is close to a 
flow generating a free circle
action (in our case, the geodesic flow on the round two-sphere), which implies that such periodic orbits are at least as many as
one plus the cup-length of $\Sf^2$, i.e. two.

\medskip

For other  available results for (\ref{P_Arnold}) showing the existence of at least two distinct  solutions for arbitrary metrics on $\Sf^2$ let us mention
\cite[Theorem 2.1 (i) and Theorem 2.7]{G2}, \cite{S0}, \cite{RS}. Notice that all these results require that $K$ has constant sign: indeed, in \cite{G2} the assumption $K>0$ guarantees that 
$\Omega= K d\sigma$ is a symplectic form on $\Sf^2$; in \cite{S0}, \cite{RS} an index-based  topological argument is used to
prove the existence of  two distinct solutions for any $c>0$, and the assumption $K>0$ is needed to prove some
crucial {\em a-priori} bound on the length of simple and closed $K$-magnetic geodesics. Schneider's multiplicity result is indeed sharp, that is, Problem (\ref{P_Arnold}) might have exactly two distinct solutions, see \cite[Theorem 1.3]{S0}.

Let us however notice that from the physical point of view it is important to include sign-changing functions $K$, unless the existence of magnetic monopoles is admitted. 
In fact, the Gauss law for magnetism in absence of magnetic monopoles implies that
$$
\int\limits_{\Sf^2} K(p)\ms d\sigma_p=0\ms,
$$
see also \cite[Problem 1996-17]{A1}.

\bigskip

The aim of this paper is twofold. Firstly, we 
provide a more direct, self-contained and analytical approach to Viterbo's and Bottkoll's results, in the special case of the round sphere.
Secondly, we provide sufficient conditions on $K$ to obtain as many solutions as we wish, provided that $c$ is large enough.

Our main results are stated in Sections \ref{S:two} and \ref{S:many}, see Theorems \ref{T:main} and \ref{T:main1}, respectively.

For the proofs we took inspiration from the  breakthrough paper \cite{AB}, where Ambrosetti and Badiale
showed  how merging the Lyapunov-Schmidt finite-dimensional reduction with variational arguments allows to obtain
extremely powerful tools to get existence and multiplicity results. This idea has been applied to tackle quite a large number of variational problems arising 
from mathematical physics and differential geometry, see the exhaustive list of references in the monograph \cite{AM}.

Notice however that  Arnol'd  problem on $K$-magnetic geodesics in 
$\Sf^2$ does not admit a (standard) variational formulation through a (non-multivalued) energy functional,
due to obvious topological obstructions.
To overcome this difficulty, we take advantage 
of a ''local'' variational approach which is developed in Section \ref{S:variational}.

\bigskip
{\small 

\noindent
{\bf Notation.}\\
The Euclidean space $\R^3$ is endowed with Euclidean  norm $|p|$,  scalar product
$p\cdot q$,  and exterior product $p\wedge q$.
The canonical basis of $\R^3$ is $\{e_h~,~h=1,2,3\}$. 

We isometrically embed the
unit sphere $\Sf^2$ into $\R^3$, so that the tangent space 
$T_z\Sf^2$ at $z\in\Sf^2$ is identified with 
$\langle z\rangle^\perp=\{p\in\R^3~|~p\cdot z=0\}$.
We denote by ${\mathcal D}_{\!\rho}(z)\subset\Sf^2$ 
the geodesic disk of radius $\rho\in(0,\frac\pi2]$ about $z\in\Sf^2$.

It is convenient to regard at $\Sf^1$ as the unit circle in the complex plane.

\medskip
\noindent
{\bf Function spaces.}~
Let $m\ge 0$, $n\ge 1$ be integer numbers. We endow  $C^m(\Sf^1,\R^n)$ with the standard Banach space structure. If $f\in C^1(\Sf^1,\R^n)$, 
we identify $f'(x)\equiv f'(x)(ix)$, so that $f':\Sf^1\to\R^n$.

We  write 
$C^m(\Sf^1)$ instead of $C^m(\Sf^1,\R)$ and $C^m$ instead of $C^m(\Sf^1,\R^3)$. For $U\subseteq\Sf^2$ we put
$$
C^m_{U}:=C^m(\Sf^1,U)=
\{u\in C^m~|~ u(x)\in U~~\text{for any $x\in\Sf^1$}\}\ms.
$$
We identify 
$U$ with the set of constant functions in 
$C^2_{U}$, so that 
$C^2_{U}\setminus U=C^2_{U}\setminus\Sf^2$ contains only nonconstant curves. 

The Hilbertian norm in 
$L^2=L^2(\Sf^1,\R^3)$ is 
$$\displaystyle{\|u\|^2_{L^2}=\fintm |u(x)|^2~\!dx~\!=\frac{1}{2\pi} \irn|u(x)|^2~\!dx}\ms,$$
and the orthogonal to $T\subseteq C^0$ with respect to the $L^2$ scalar product is 
given by
$$
T^\perp=\{\varphi\in C^0~|~\fintm u\cdot\varphi~\!dx=0~~\text{for any $u\in T$}~\}.
$$
We regard at $C^2_{\Sf^2}$ as a smooth complete submanifold of $C^2$.
If $u\in C^2_{\Sf^2}$, the tangent space to $C^2_{\Sf^2}$ at $u$ is 
$$
T_uC^2_{\Sf^2}=\{\varphi\in C^2~|~ u\cdot \varphi\equiv 0~\text{on $\Sf^1$}~\}.
$$
If $u$ is regular, that means $u'(x)\neq 0$ for any $x\in\Sf^1$, then 
$$
T_uC^2_{\Sf^2}=\{g_1 u'+g_2 \ms u\wedge u'~|~ g=(g_1,g_2)\in C^2(\Sf^1,\R^2)~\}.
$$

\medskip
\noindent
{\bf Rotations.}
Any complex number $\xi\in\S^1$ is identified with the rotation $x\mapsto \xi x$. 
Recall that $\det(R) = +1$ and $R^{-1} = \prescript{t}{}{\!R}$
for any $R\in SO(3)$, where $SO(3)$ is the group of rotations of $\R^3$ and 
$\prescript{t}{}{\!R}$ is the transpose of $R$.

It is well-known that $SO(3)$  is a connected three-dimensional manifold. More precisely, it is a Lie group whose Lie algebra
is given by the skew-symmetric matrices, and the tangent space $T_{\Id}SO(3)$ at the identity matrix is spanned by 
$$
T_1=\left(
\begin{array}{ccc}
0&0&0\\
0&0&-1\\
0&1&0
\end{array}
\right),~
{T_2=\left(
\begin{array}{ccc}
0&0&1\\
0&0&0\\
-1&0&0
\end{array}
\right)},~
T_3=\left(
\begin{array}{ccc}
0&-1&0\\
1&0&0\\
0&0&0
\end{array}
\right)
\ms.
$$
A simple explanation of this elementary fact follows by introducing the matrices 
$$
{R}_1^\xi=\left(
\begin{array}{ccc}
1&0&0\\
0&\xi_1&-\xi_2\\
0&\xi_2&\xi_1
\end{array}
\right),~
{{R}_2^\xi=\left(
\begin{array}{ccc}
\xi_1&0&-\xi_2\\
0&1&0\\
\xi_2&0&\xi_1
\end{array}
\right)},~
{R}_3^\xi=\left(
\begin{array}{ccc}
\xi_1&-\xi_2&0\\
\xi_2&\xi_1&0\\
0&0&1
\end{array}
\right)
$$
for  $\xi=\xi_1+i\xi_2\in\Sf^1$. 
Clearly $R_h^\xi$ is a rotation about the $\langle \ms e_h\ms\rangle$ axis.
By differentiating $R_h^\xi$ with respect to $\xi\in \Sf^1$ at $\xi=1$ one gets
$T_h=d{{R}^\xi_h}_{\big|\xi=1}$, and thus infers that $\{T_h\}$ is a basis for $T_{\Id}SO(3)$. 
In accordance with the  Lie group structure of $SO(3)$, the tangent space to $SO(3)$ at $R\in SO(3)$
is obtained by rotating $T_{\Id}SO(3)$. Hence
$$
T_\mR\SO=\langle\mR T_1,\mR T_2,\mR T_3\rangle.
$$
Finally, for any $q\in\Sf^2$ we denote by $d_R$ the differential of the function $\SO\to\Sf^2$, $R\mapsto Rq$,
so that $d_R(Rq)\tau\in T_{Rq}\Sf^2$ for any $\tau\in T_R\SO$. We have the formula
\begin{equation}
\label{eq:dR}
d_R(Rq)(RT_h)=R(e_h\wedge q)=Re_h\wedge Rq.
\end{equation}}
\section{A "local" variational approach}
\label{S:variational}

We put $\eps=c^{-1}$ and study Problem ($\mathcal P_{K,\eps^{-1}}$) for $\eps$ close to $0$. We
take advantage of its geometrical interpretation to rewrite it 
in an equivalent way. Let $\gamma$ be a solution to  ($\mathcal P_{K,\eps^{-1}}$), and let $\mathcal L_\gamma$ be its length. Extend $\gamma$
to an $\eps\mathcal L_\gamma$-periodic function on $\R$ and consider the curve
$u\in C^2_{\Sf^2}$, $u(e^{i\theta})=\gamma\big(\frac{\eps\mathcal L_\gamma}{2\pi}\ms\theta\big)$.
Evidently $u$ and $\gamma$ have the same length $\mathcal L_\gamma$ and 
curvature $\eps K$. Moreover 
$|u'|\equiv \mathcal L_\gamma/2\pi$ and $u$ solves the system
\begin{equation}\label{eq:EL}
u''+|u'|^2u=|u'|\eps K(u)~\!u\wedge u'~\qquad \text{on $\Sf^1$},
\end{equation}
because $\gamma$ solves (\ref{eq:Arnold}).
Conversely, any solution $u\in C^2_{\Sf^2}\setminus\Sf^2$ to (\ref{eq:EL}) has constant speed $|u'|$,
curvature $\eps K(u)$ and gives rise to a solution to ($\mathcal P_{K,\eps^{-1}}$). 

The main goal of the present section is to show that for any point $p\in\Sf^2$, the problem
of finding solutions to (\ref{eq:EL}) in $C^2_{\Sf^2\setminus\{p\}}$, that is an open subset of
$C^2_{\Sf^2}$, can be faced by using variational methods.
First, we need to introduce the functional 
\begin{equation}
\label{eq:L}
L(u)=\Big(\fintm |u'|^2~\!dx\Big)^\frac{1}{2}~,\quad
L:C^2_{\Sf^2}\setminus\Sf^2\to\R.
\end{equation}
Notice that the Cauchy-Schwarz inequality gives
$\displaystyle{\mathcal L_u\le 2\pi L(u)}$,
and equality holds if and only if $|u'|$ is constant. Moreover, it holds that 
\begin{equation}
\label{eq:L_invariance}
L({R} u\circ\xi)=L(u)\quad\text{for any $\xi\in \Sf^1,~{R}\in SO(3)$.}
\end{equation}
Finally, we notice that $L$ is Fr\'echet differentiable at any $u\in C^2_{\Sf^2}\setminus\Sf^2$, with differential
\begin{equation}
L'(u)\varphi
=\frac{1}{L(u)}~\!\fintm u'\cdot\ \varphi'\ms dx
=\frac{1}{L(u)}~\!\fintm (-u''-|u'|^2u)\cdot\varphi \ms dx
\quad \text{for any $\varphi\in
T_uC^2_{\Sf^2}$.}
\label{eq:dL}
\end{equation}
In the next lemma we provide a variational reading of the right-hand side of (\ref{eq:EL}),
see also  \cite{NT} and \cite[Remark 2.2]{G2}.

\begin{Lemma}
\label{L:area0}
Let $K \in C^0(\Sf^2)$ and let $U, V$ be open and contractible subsets of $\Sf^2$.
\begin{itemize}
\item[$i)$] 
There exists a unique $C^1$ functional 
$\mathcal A^U_K: C^2_U \to \R$, such that $\mathcal A^U_K( u) = 0$ if $u$ is constant, and
\begin{equation}
\label{eq:dA0}
(\mathcal A^U_K)'( u) \phi = \fint_{S^1} K(u) \phi \cdot u \wedge u' \ms dx\quad
\text{for any $u\in C^2_U$, $\phi \in T_u C^2_{\Sf^2}$;}
\end{equation}
\item[$ii)$] If $R\in SO(3)$, $\xi\in \Sf^1$ and 
$u\in C^2_U$, then ${\mathcal A^{RU}_{K\circ \prescript{t}{}{\!R}}(Ru\circ \xi)=\mathcal A^U_K(u)}$;
\item[$iii)$] If $U\cap V$ is nonempty and contractible,  then
$\mathcal A^U_K(u)=\mathcal A^V_K(u)$ for any $u\in C^2_{U\cap V}$;
\item[$iv)$] Let $u\in C^2_{\Sf^2}$. The function 
$p \mapsto \mathcal A_K^{\Sf^2\setminus\{p\}}(u)$ is constant on each connected component of $\Sf^2 \setminus u(\Sf^1)$;
\item[$v)$] Let $u\in C^2_U$ be a positively oriented parametrization of the boundary 
of a regular open set $\Omega_u \subset U$. Then
$$
\mathcal A_K^U(u) = -\frac{1}{2 \pi} \int\limits_{\Omega_u} K(q)\ms d\sigma_{\!q}\ms.$$
\end{itemize}
\end{Lemma}

\proof
Take a $1$-form $\beta_K^U$ on  $U$, such that 
\begin{equation}
\label{eq:dbeta}
d \beta_K^U ={-}  K(q) \ms d\sigma_{\!q}\ms,
\end{equation}
where
$d\sigma_{\!q}$ is the restriction of the volume form on the sphere. We put 
$$
\mathcal A^U_K( u) = \fintm u^*\beta_K^U = \fintm \beta_K^U(u) u'~\! dx~,\quad u\in C^2_U\ms.
$$
It is evident that $\mathcal A^U_K( u)=0$ if $u$ is constant. Formula (\ref{eq:dA0})
can be derived by using Lie differential calculus or local coordinates, like
in the proof of \cite[Lemma 3]{BT}.
Elementary arguments and (\ref{eq:dA0}) give the $C^1$ differentiability of the functional
$\mathcal A^U_K$. 
Uniqueness is trivial, because $C^2_U$ is a connected manifold. 
In particular, for $u\in C^2_U$ the real number  $\mathcal A^U_K(u)$
does not depend on the choice of $\beta_K^U$. 

\medskip
To prove $ii)$ take a $1$-form $\beta$ in the domain $RU$ such that $d\beta=-(K\circ \prescript{t}{}{\!R}) \ms d\sigma_{\!q}$.
Clearly  $R^*\beta$ is a $1$-form in $U$,
and $d(R^*\beta)=R^*(d\beta)= 
- K(q)d\sigma_{\!q}$. Thus we can take $\beta_K^U=R^*\beta$ in formula (\ref{eq:dbeta}) and we obtain
$${\mathcal A^{RU}_{K\circ \prescript{t}{}{\!R}}(Ru)= \fintm (Ru)^*\beta = \fintm u^* (R^*\beta)  = \mathcal A^U_K(u)}$$ for any $u\in C^2_U$. The invariance
of the area functional with respect to composition with rotations of $\Sf^1$ is immediate.

\medskip

Now we prove $iii)$. If $V\subset U$ and $u\in C^2_V$, then the restriction of $\beta^U_K$ to $V$
can be used to compute $\mathcal A_K^V(u)$. Thus $\mathcal A_K^V(u)=\mathcal A_K^U(u)$.
It follows that if two open, connected sets $U,V$ have contractible intersection and $u\in C^2_{U\cap V}$,
then $\mathcal A_K^{U\cap V}(u)=\mathcal A_K^U(u)$ and $\mathcal A_K^{U\cap V}(u)=\mathcal A_K^V(u)$.

\medskip

Claim $iv)$ readily follows from $iii)$. In fact, take $p_0\in \Sf^2\setminus u(\Sf^1)$ and a small 
disk $\mathcal D_\delta(p_0)\subset \Sf^2\setminus u(\Sf^1)$.
For any $p\in \mathcal D_\delta(p_0)$ we have
$$
\mathcal A^{\Sf^2\setminus\{p\}}(u)=\mathcal A^{\Sf^2\setminus \mathcal D_\delta(p_0)}(u)=
\mathcal A^{\Sf^2\setminus\{p_0\}}(u)\ms.
$$
We proved that the function $p\mapsto \mathcal A^{\Sf^2\setminus\{p\}}(u)$ is locally constant on $\Sf^2\setminus u(\Sf^1)$, and hence is constant on each connected component of $\Sf^2 \setminus u(\Sf^1)$.

For the last claim we 
use Stokes' theorem to get 
$$
2\pi \mathcal A_K^U(u) = \int\limits_{\Sf^1} u^*\beta^U_K =\int\limits_{\partial\Omega_u} \beta^U_K= 
\int\limits_{\Omega_u} d \beta^U_K = -\int\limits_{\Omega_u} K(q) d\sigma_{\!q} 
$$
by (\ref{eq:dbeta}).
The lemma is completely proved.
\QED

From now on we write
$$
A_K(p;u)= \mathcal A_K^{\Sf^2\setminus\{p\}}(u) ~,\quad p\in\Sf^2~,~~ u\in C^2_{\Sf^2\setminus\{p\}}.
$$
By Lemma \ref{L:area0}, the functional $A_K$ enjoys the following properties,
\begin{itemize}
\item[$A1)$] The functional $A_K(p; \cdot)$ is of class $C^1$ on $C^2_{\Sf^2\setminus\{p\}}$, and
$$
A_K'(p; u) \phi = \fint_{S^1} K(u) \phi \cdot u \wedge u' \ dx \ \ 
\text{for any $u\in C^2_{\Sf^2\setminus\{p\}}$, $\phi \in T_u C^2_{\Sf^2}$.}
$$
\item[$A2)$] If $R\in SO(3)$, $\xi\in \Sf^1$, and 
$u\in C^2_{\Sf^2\setminus\{p\}}$, then ${A_{K\circ \prescript{t}{}{\!R}}(Rp;Ru\circ \xi)=A_K(p;u)}$.
\item[$A3)$] Let $u \in C^2_{\Sf^2}$. The function $p \mapsto A_K(p; u)$ is locally constant on $\Sf^2 \setminus u(\Sf^1)$.
\item[$A4)$] Let $u\in C^2_{\Sf^2\setminus\{p\}}$ be a positively oriented parametrization of the boundary of a regular open set $\Omega_u \subset \Sf^2 \setminus \{p\}$. Then
$$A_K(p; u) = -\frac{1}{2 \pi} \int\limits_{\Omega_u} K(q)\ms d\sigma_{\!q}\ms.$$
\end{itemize}

\begin{Remark}
\label{R:A_explicit}
To find an explicit formula for $A_K(p;\ms\cdot\ms)$ let $\Pi_p:\Sf^2\setminus\{p\}\to \R^2$ be the stereographic projection 
from the pole $p$. If $u\in C^2_{\Sf^2\setminus\{p\}}$, then $\Pi_p\circ u$ is a curve in 
$\R^2$ and  $(\Pi_p^{-1})^*(Kd\sigma_{\!q})=(K\circ \Pi_p^{-1}){\rm det}J_{\Pi_p^{-1}}(z)dz$ 
is a $2$-form on $\R^2$. Let
$\tilde \beta_K^p$ be a $1$-form on $\R^2$ such that $d\tilde \beta_K^p=(\Pi_p^{-1})^*(Kd\sigma_{\!q})$. Then 
$$
A_K(p;u)=\fintm u^*(\Pi_p^*\ms \tilde \beta_K^p)\ms=\fintm(\Pi_p\circ u)^*\tilde \beta_K^p\ms.
$$
For instance, if  $K\equiv 1$ is constant one can take
$$
A_1(p;u) 
=\fintm \frac{p}{1-u\cdot p}~\cdot u\wedge u'\ms dx
=2\fintm \frac{p}{|u-p|^2}\cdot u\wedge u'\ms dx.
$$
\end{Remark}
The next lemma provides the predicted "local" variational approach to (\ref{eq:EL}).

\begin{Lemma}
\label{L:E}
Let $K\in C^0(\Sf^2)$.

$i)$  For any $p\in\Sf^2$, the functional 
$$
E_{\eps K}(p;u)=L(u)+\eps A_K(p;u)\ms, \quad E_{\eps K}(p;\ms\cdot\ms): C^2_{\Sf^2\setminus\{p\}}\setminus\Sf^2\to \R
$$
is of class $C^1$, with differential
\begin{equation}
\label{eq:per_EJ}
L(u)E'_{\eps K}(p;u)\varphi
=\fintm \big(-u''+ L(u)~\!\eps K(u) u\wedge u'\big)\cdot\varphi\ms dx,\quad 
\text{for any $\varphi\in T_u C^2_{\Sf^2}$.}
\end{equation}
In particular, any  critical point $u\in C^2_{\Sf^2\setminus\{p\}}\setminus\Sf^2$  for 
$E_{\eps K}(p;\ms\cdot\ms)$ solves {\rm (\ref{eq:EL})}.

$ii)$ If $R\in SO(3)$, $\xi\in \Sf^1$ and $p\in\Sf^2$, then
$E_{{\eps K}\circ \prescript{t}{}{\!R}}(Rp;Ru\circ \xi)=E_{\eps K}(p;u)$ for any nonconstant curve $u\in C^2_{\Sf^2\setminus\{p\}}$, and thus
\begin{equation}
\label{eq:EK_invariance}
E'_{\eps K}(p;u)u'=0\quad \text{for any $u\in C^2_{\Sf^2\setminus\{p\}}\setminus\Sf^2$.}
\end{equation}

$iii)$ Let $u\in C^2_{\Sf^2}\setminus\Sf^2$. The function 
$E_{\eps K}(\ms\cdot\ms;u):\Sf^2\setminus u(\Sf^1)\to\R$ is locally constant.

$iv)$
If $K\in C^1(\Sf^2)$ then the functional $E_{\eps K}(p;\ms\cdot\ms)$ is of class $C^2$ on its domain.
\end{Lemma}

\proof
Formula (\ref{eq:dL}) and the property $A1)$ of the area functional give the $C^1$ regularity of $E_{\eps K}(p;\ms\cdot\ms)$ and 
(\ref{eq:per_EJ}).  Let $u$ be a critical point for 
$E_{\eps K}(p;\ms\cdot\ms)$. Take any $\varphi\in C^2$
and put $\varphi^\top=\varphi-(\varphi\cdot u)u\in T_u C^2_{\Sf^2}$.
 We  have
$\varphi\cdot u\wedge u'= \varphi^\top\cdot u\wedge u'$ on $\Sf^1$, and
$u'\cdot(\varphi^\top)'=u'\cdot\varphi'-(\varphi\cdot u)|u'|^2$
because $u'\cdot u\equiv 0$. Since 
$$
\begin{aligned}
0=L(u)E'_{\eps K}(p;u)\varphi^\top=&
\fintm \big(u'\cdot(\varphi^\top)'+L(u)~\!\eps K(u)\varphi^\top\cdot u\wedge u'\big)\ms dx\\
=&\fintm \big(u'\cdot\varphi'-(\varphi\cdot u)|u'|^2+L(u)~\!\eps K(u)\varphi\cdot u\wedge u'\big)\ms dx\ms,
\end{aligned}
$$
and therefore $u$ solves 
$u''+|u'|^2u=L(u)~\!\eps K(u)~\!u\wedge u'$ on  $\Sf^1$. Since $u''\cdot u'\equiv 0$, we see that $|u'|\equiv L(u)$ is constant, and thus $u$ solves (\ref{eq:EL}).

Statements $ii)$, $iii)$ follow from (\ref{eq:L_invariance}), $A2)$ and $A3)$ (to check (\ref{eq:EK_invariance}) take
the derivative of the identity
$E_{\eps K}(p;u\circ \xi)=E_{\eps K}(p;u)$ with respect to $\xi\in\Sf^1$ at $\xi=1$). Finally, $iv)$  can be proved via 
elementary arguments, starting from (\ref{eq:per_EJ}).
\QED

\section{Geodesics}
\label{S:constant}

For any rotation $R\in \SO$, the loop
$$
\omega_{\!R}(x)=R\big(x_1,x_2,0)~,\quad x=x_1+ix_2\in\Sf^1~\!,
$$
is a parameterization of the boundary of ${\mathcal D}_{\!{\frac\pi2}}(Re_3)$ and
solves
\begin{equation}
\label{eq:kequ}
\omega_{\!R}''+|\omega_{\!R}'|^2\omega_{\!R}=0~,\quad L(\omega_{\!R})=|\omega_{\!R}'|={1}\ms.
\end{equation}
In order to simplify notations, from now on we write 
$$
\omega(x)=\omega_{\rm Id}(x)=\big(x_1,x_2,0)~,\quad x=x_1+ix_2\in\Sf^1~\!.
$$
The tangent space to the smooth $3$-dimensional manifold
$$
\mathcal S=\big\{\omega_{\!R}~|~{R}\in SO(3)\ms\big\}\subset C^2_{\Sf^2}
$$
at $\omega_{\!R}\in \mathcal S$ can be easily computed via formula (\ref{eq:dR}). It turns out that 
$$
T_{\omega_{\!R}}\mathcal S =\{q\wedge \omega_{\!R}~|~ q\in\R^3\}=
\langle {R}e_1\wedge \omega_{\!R}~,~
Re_2\wedge \omega_{\!R}~,~Re_3\wedge \omega_{\!R}~\rangle.
$$
We introduce the function 
$$
J_0(u):= -u''-|u'|^2u~,\quad J_0:C^2_{\Sf^2}\setminus \Sf^2\to C^0,
$$
so that $\mathcal S\subset\{J_0=0\}$. By  (\ref{eq:dL}) we have
\begin{equation}
\label{eq:EJ}
L(u)L'(u)\varphi=\fintm J_0(u)\cdot\varphi\ms dx\quad
\text{for any $u\in C^2_{\Sf^2}\setminus\Sf^2$, $\varphi\in T_uC^2_{\Sf^2}$.}
\end{equation}
Moreover, for $u\in C^2_{\Sf^2}\setminus\Sf^2$, $q\in\R^3$ and ${R}\in SO(3)$ it holds that
\begin{equation}
\label{eq:inv0}
\fintm J_0(u)\cdot q\wedge u~\!dx=0~,\quad
J_0({R} u)=RJ_0(u)\ms.
\end{equation}
The first identity  readily
follows via integration by parts or can be obtained by 
differentiating  the identity $L(Ru)=L(u)$ with respect to $R\in SO(3)$.
The second one is immediate. 

Clearly $J_0$ is of class $C^2$; 
for $R\in SO(3)$ and $\varphi$ in the tangent space
\begin{equation}
\label{eq:tan_frame}
T_{\omega_{\!R}}C^2_{\Sf^2}=\{\varphi=g_1\ms \omega_{\!R}'+g_2
\ms \omega_{\!R}\wedge \omega_{\!R}'~|~g=(g_1,g_2)\in C^2(\Sf^1,\R^2)~\},
\end{equation}
we have
$$
J_0'(\omega_{\!R})\varphi=-\varphi''-2(\omega_{\!R}'\cdot\varphi')\omega_{\!R}-\varphi\\
\ms.
$$
Further, the operator $J'_0(\omega_R)$ is self adjoint in $L^2(\Sf^1,\R^3)$, that is, 
\begin{equation}
\label{eq:self}
\fintm J'_0(\omega_R)\varphi\cdot{\tilde\varphi}\ms dx=\fintm J'_0(\omega_R){\tilde\varphi}\cdot\varphi\ms dx
\quad\text{for any $\varphi,\tilde\varphi\in T_{\omega_R}C^2_{\Sf^2}$.}
\end{equation}
By differentiating the identity $J_0(\omega_{\!R})=0$ with respect to
${R}\in SO(3)$, we see that $T_{\omega_{\!R}}\mathcal S\subseteq \ker\! J_0'(\omega_{\!R})$.
Actually, equality holds, as shown in the next crucial lemma.

\begin{Lemma}[Nondegeneracy]
\label{L:nondegen}
Let ${R}\in SO(3)$. Then 
\begin{itemize}
\item[$i)$] $\displaystyle{\ker\! J'_0(\omega_{\!R})=T_{\omega_{\!R}}\mathcal S}$;

\item[$ii)$] If $\varphi\in T_{\omega_{\!R}}C^2_{\Sf^2}$ and $J_0'(\omega_{\!R})\varphi\in T_{\omega_{\!R}}\mathcal S$, then 
$\varphi\in T_{\omega_{\!R}}\mathcal S$;

\item[$iii)$] 
For any $u\in T_{\omega_{\!R}}\mathcal S^\perp$
there exists a unique $\varphi\in {T_{\omega_{\!R}}C^2_{\Sf^2}}\cap T_{\omega_{\!R}}\mathcal S^\perp$ such that $J_0'(\omega_{\!R})\varphi=u$. 
\end{itemize}
\end{Lemma}

\proof
One can argue by adapting the computations in \cite[Section 5]{S0}. 
We provide here a simpler argument.

Since $J'_0(\omega_{\!R})(R\varphi)=R\big(J'_0(\omega)\varphi\big)$ for any $\varphi\in T_{\omega}C^2_{\Sf^2}$,
it is not restrictive to assume that $R$ is the identity matrix.
By direct computations based on (\ref{eq:kequ}), one can check that 
$$
J'_0(\omega)(\psi\ms  \omega')=-\psi''\ms \omega'~, \quad
J'_0(\omega)(\psi \ms \omega\wedge \omega')=
\big(-\psi'' -\psi\big)\ms \omega\wedge \omega'
$$
for any  $\psi\in C^2(\Sf^1,\R)$. Since by (\ref{eq:tan_frame}) any
function $\varphi\in T_{\omega}C^2_{\Sf^2}$ can be written
as 
$$
\varphi=(\varphi\cdot\omega')\omega'+(\varphi\cdot \omega\wedge\omega')\ms \omega\wedge\omega'\ms,
$$
we
are led to 
introduce the differential operator $B:C^2(\Sf^1,\R^2)\to C^0(\Sf^1,\R^2)$,
$$
B(g)=
-g_1''\ms e_1+(-g_2''-g_2)e_2~,\qquad g=(g_1,g_2)\in C^2(\Sf^1,\R^2)\ms.
$$
and the function transform 
$$
\Psi\varphi=(\varphi\cdot\omega')\ms e_1+\ms 
(\varphi\cdot\omega\wedge\omega')\ms e_2\ms,
\quad \Psi: T_{\omega}C^2_{\Sf^2}\to C^2(\Sf^1,\R^2)\ms,
$$
so that
\begin{equation}
\label{eq:formula}
J_0'(\omega)\varphi=\Psi^{-1}B(\Psi\varphi)\quad\text{for any $\varphi\in T_{\omega}C^2_{\Sf^2}$,}\quad
\Psi(\ker \! J'_0(\omega))=\ker B\ms.
\end{equation}
We proved that $\ker\! J'_0(\omega)$ and $T_{\omega}\mathcal S$ have both dimension $3$, thus they must coincide
because $T_{\omega}\mathcal S\subseteq\ker\! J'_0(\omega)$.

For future convenience we notice that $\Psi$ is an isometry with respect to the $L^2$ norms, and in particular 
\begin{equation}
\label{eq:L2inv}
\fintm\big(\Psi\varphi\big)\cdot \big(\Psi\tilde\varphi\big)\ms dx=
\fintm\varphi\cdot\tilde\varphi\ms dx\quad \text{for any $\varphi,\tilde\varphi\in T_{\omega}C^2_{\Sf^2}$. }
\end{equation}

Now we prove $ii)$. If $\tau:=J_0'(\omega)\varphi\in T_{\omega}\mathcal S$,
then $J_0'(\omega)\tau=0$, as $\displaystyle{\ker\! J_0'(\omega)=T_{\omega}\mathcal S}$.
But then, using (\ref{eq:self}) we get
$$
\fintm|J_0'(\omega)\varphi|^2~\!dx=\fintm J_0'(\omega)\varphi\cdot\tau~\!dx=\fintm J_0'(\omega)\tau\cdot\varphi~\!dx=0.
$$
Thus $J_0'(\omega)\varphi=0$, that means $\varphi \in T_{\omega}\mathcal S$. 

It remains to prove $iii)$. Since $\Psi(T_\omega\mathcal S)=\ker B$, from (\ref{eq:formula}) and
(\ref{eq:L2inv}) we have that 
$u \in T_{\omega}\mathcal S^\perp$  if and only if $\Psi u\in \ker B^\perp$.
In particular, if  $u\in T_{\omega}\mathcal S^\perp$, then
one can  compute the unique solution $g_u\in \ker B^\perp$
 to the system $Bg_u=\Psi u$. 
The function $\varphi:=\Psi^{-1}g_u$ belongs to $T_{\omega}\mathcal S^\perp$;
thanks to (\ref{eq:formula}) it
solves
$J_0'(\omega)\varphi=u$, and is uniquely determined by $u$.
The lemma is completely proved.
\QED

\begin{Remark}
\label{R:M}
For future convenience we compute 
$$
m_{hj}=\fintm (Re_h\wedge \omega_{\!R})\cdot (Re_j\wedge \omega_{\!R})\ms dx=
\fintm (e_h\wedge \omega)\cdot (e_j\wedge \omega)\ms dx=
\delta_{hj}-\fintm\omega_h\omega_j~\!dx.
$$
We see that the functions ${R}e_j\wedge \omega_{\!R}=R(e_j\wedge\omega)$
provide an orthogonal basis for $T_{\omega_{\!R}}\mathcal S$ endowed
with the $L^2$ scalar product. More precisely, the 
matrix $M$ associated to this scalar product with respect to the basis $\{{R}e_j\wedge \omega_{\!R}\}$
is given by
$$
M=\left(\begin{array}{ccc}
\frac12&0&0\\
0&\frac12&0\\
0&0&1
\end{array}\right)\ms.
$$
\end{Remark}

\subsection{Finite dimensional reduction}
\label{SS:prova}

By the remarks at the beginning of Section \ref{S:variational}, we are led to 
study problem (\ref{eq:EL}) for $\eps=c^{-1}$ close to $0$. Further, since any solution $u$
to (\ref{eq:EL}) satisfies $|u'|\equiv L(u)$, we can rewrite
 (\ref{eq:EL}) in the following, equivalent way,
\begin{equation}
\label{eq:kepseq}
u''+|u'|^2u={L(u)}\eps K(u)~\! u\wedge u'~,\qquad u\in C^2_{\Sf^2}\setminus\Sf^2\ms.
\end{equation}
We will look for
solutions to (\ref{eq:kepseq}) by solving $J_\eps(u)=0$, where
$J_\eps: C^2_{\Sf^2}\setminus\Sf^2\to C^0$,
\begin{equation}
\label{eq:Jeps0}
J_\eps(u)=J_0(u)+\eps L(u)K(u)~\!u\wedge u'
=-u''-|u'|^2u+L(u)\eps K(u)~\! u\wedge u'.
\end{equation}
Thanks to (\ref{eq:per_EJ}), we can write 
\begin{equation}
\label{eq:EJeps}
L(u)E'_{\eps K}(p;u)\varphi=\fintm J_\eps(u)\cdot\varphi\ms dx~,\quad
\text{for $u\in C^2_{\Sf^2}\setminus\Sf^2$, $p\notin u(\Sf^1)$, $\varphi\in T_uC^2_{\Sf^2}$.}
\end{equation}
The regularity assumption on $K$ implies that 
$J_\eps$ is of class $C^1$ on its domain. In addition,
$J_\eps(u\circ\xi)=J_\eps(u)$ for any $\xi\in\Sf^1$, and integration by parts gives
$$
\fintm J_\eps(u)\cdot u'~\!dx=0
\qquad\text{for any $u\in C^2_{\Sf^2}\setminus\Sf^2$.}
$$
In general, the identities in (\ref{eq:inv0})
are not satisfied if $\eps\neq 0$, because the perturbation term breaks  the invariances of the operator
$J_0$. 

In the next lemma we provide the main step to obtain our multiplicity results.

\begin{Lemma}
\label{L:reduction} 
There exist $\beps>0$ and a $C^1$ function
$$
[-\beps,\beps]\times SO(3)\to C^2_{\Sf^2}\setminus\Sf^2~\quad (\eps,{R})\mapsto u^\eps_{R}
$$
such that $u^\eps_{R}$ is an embedded loop,
and moreover
\begin{itemize}
\item[$i)$] $u^0_{R}=\omega_{\!R}$;
\item[$ii)$] $u^\eps_{R}\in T_{\omega_{\!R}}\mathcal S^\perp$;
\item[$iii)$] $J_\eps(u^\eps_{R}) \in T_{\omega_{\!R}}\mathcal S$;
\item[$iv)$] The function
$[-\beps,\beps]\times SO(3)\to \R$, 
$$(\eps,{R})\mapsto \mathcal E^\eps({R}):=E_{\eps K}(-Re_3;u^\eps_R)=
L(u^\eps_R)+\eps A_{K}(-Re_3;u^\eps_R)
$$
is well defined, of class $C^1$ on its domain, and
$d_R\mathcal E^\eps({R})(RT_3)=0$.
\item[$v)$] $R\in SO(3)$ is critical for $\mathcal E^\eps: SO(3)\to\R$ if and only if
$J_\eps(u^\eps_{R})=0$.
\item[$vi)$] Put $\mathcal E^\eps_0({R})=E_{\eps K}(-Re_3;\omega_{\!R})=1+\eps A_{K}(-Re_3,\omega_{\!R})$.
As $\eps\to 0$, we have
\begin{equation}
\label{eq:zero_order}
\mathcal E^\eps({R})- \mathcal E^\eps_0({R})=o(\eps)
\end{equation}
uniformly on $SO(3)$, together with the derivatives with respect to $R\in SO(3)$.
\end{itemize}
\end{Lemma}

\proof
Consider the differentiable functions 
\begin{gather*}
\displaystyle{\mathcal F_1:\R\!\times\! SO(3)\!\times\! (C^2_{\Sf^2}\!\!\setminus\!\Sf^2)\!\!\times\!\!\R^3\to C^0}~,~
\displaystyle{\mathcal F_1(\eps,{R},u;\zeta)= 
J_\eps(u)- \sum_{j=1}^3 \zeta_j\ms (Re_j\wedge \omega_{\!R})} \\
\displaystyle{
\mathcal F_2:\R\!\times\! SO(3)\!\times\! (C^2_{\Sf^2}\!\!\setminus\!\Sf^2)\!\!\times\!\!\R^3\to\R^3}~,~
{\mathcal F_2(\eps,{R},u;\zeta)=
\sum_{j=1}^3\!\big(\displaystyle{\fintm u\cdot Re_j\wedge \omega_{\!R}\ms  dx\big)e_j}}
\end{gather*}
where $\zeta=(\zeta_1,\zeta_2,\zeta_3)\in\R^3$,
and then let 
$$
\mathcal F:\R\times SO(3)\times (C^2_{\Sf^2}\!\!\setminus\!\Sf^2)\!\times\!\R^3\to C^0\!\times\! \R^3~,\quad
\mathcal F=\big(\mathcal F_1,\mathcal F_2).
$$

\smallskip

\noindent
Fix ${R}\in SO(3)$. Since $J_0(\omega_{\!R})=0$ by (\ref{eq:kequ}), then
$\mathcal F(0,{R},\omega_{\!R};0)=0$.
Our first goal is to solve the equation $\mathcal F(\eps,{R},u;\zeta)=(0;0)$ 
in a neighborhood of $(0,{R},\omega_{\!R};0)$,
via the implicit function theorem.

Consider the differentiable function 
$$
\mathcal F(0,{R},\ms\cdot\ms;\ms\cdot\ms) \ms:\ms (u;\zeta)\mapsto \mathcal F(0,{R},u;\zeta)\ms,
\quad (C^2_{\Sf^2}\!\!\setminus\!\Sf^2)\!\times\!\R^3\to C^0\!\times\! \R^3
$$
and let 
$$
\mathcal L=(\mathcal L_1,\mathcal L_2):(T_{\omega_{\!R}}C^2_{\Sf^2})\!\times\!\R^3\to C^0\!\times\!\R^3$$
be its differential evaluated at 
$(u;\zeta)=(\omega_{\!R};0)$. We need to prove that $\mathcal L$ is invertible.

Take $\varphi\in T_{\omega_{\!R}}C^2_{\Sf^2}$ and $p=(p_1,p_2,p_3)\in \R^3$. It is easy to compute
$$
\mathcal L_1(\varphi;{p})= 
J_0'(\omega_{\!R}) \varphi-\sum_{j=1}^3 \ms p_j\ms (Re_j\wedge \omega_{\!R}),
\quad
\mathcal L_2(\varphi;{p})=\sum_{j=1}^3
\big( \fintm  \varphi\cdot{R}e_j\wedge \omega_{\!R}~\!dx\big)e_j\ms.
$$
Next, recall that $T_{\omega_{\!R}}\mathcal S$ is spanned by the functions $Re_j\wedge \omega_{\!R}$.
If
$\mathcal L_1(\varphi;{p})=0$ then $J_0'(\omega_{\!R}) \varphi\in T_{\omega_{\!R}}\mathcal S$, hence 
$ \varphi\in T_{\omega_{\!R}}\mathcal S$ by $ii)$ in Lemma \ref{L:nondegen}; if $\mathcal L_2(\varphi;{p})=0$
then $ \varphi\in T_{\omega_{\!R}}\mathcal S^\perp$. Therefore, the operator $\mathcal L$ is 
injective. 

\medskip

Before proving surjectivity we notice that 
\begin{equation}
\label{eq:Ort}
J'_0(\omega_{\!R})\varphi\in T_{\omega_{\!R}}\mathcal S^\perp\quad\text{for any $\varphi\in T_{\omega_{\!R}}C^2_{\Sf^2}$}
\end{equation}
because of (\ref{eq:self}) and since $T_{\omega_{\!R}}\mathcal S=\ker \! J'_0(\omega_{\!R})$.

Now take arbitrary $\psi\in C^0$ and ${q}=(q_1,q_2,q_3)\in\R^3$. We have to find functions
$\varphi^\top\in T_{\omega_{\!R}}\mathcal S, \varphi^\perp\in T_{\omega_{\!R}}\mathcal S^\perp$ and ${p}=(p_1,p_2,p_3)\in\R^3$ such that 
$\mathcal L(\varphi^\top+\varphi^\perp,p)=(\psi,q)$. Since $T_{\omega_{\!R}}\mathcal S=\ker \! J'_0(\omega_{\!R})$ is spanned by the functions
$Re_j\wedge \omega_{\!R}$,
we only need to solve
$$
\begin{cases}
J_0'(\omega_{\!R})\varphi^\perp=\psi+\sum_{j}\ms  p_j ({R}e_j\wedge \omega_{\!R}),&
\varphi^\perp\in T_{\omega_{\!R}}\mathcal S,~p\in\R^3\\
\displaystyle\fintm\varphi^\top\cdot Re_j\wedge \omega_{\!R}~\!dx=q_j,&
 \varphi^\top\in T_{\omega_{\!R}}\mathcal S^\perp.
\end{cases}
$$
The tangential component $\varphi^\top\in T_{\omega_{\!R}}\mathcal S$  is uniquely determined.
Thanks to (\ref{eq:Ort}), we see that the function $\sum_{j}\ms  p_j ({R}e_j\wedge \omega_{\!R})$
must coincide with the projection of $-\psi$ on $T_{\omega_{\!R}}\mathcal S$. This gives the unknown $p$. More explicitly, we have
$$
e_h\cdot Mp = \sum_{j=1}^3 p_j\fintm (Re_h\wedge \omega_{\!R})\cdot (Re_j\wedge \omega_{\!R})\ms dx
=
-\fintm\psi\cdot Re_h\wedge \omega_{\!R}\ms dx\ms,
$$
where $M$ is the invertible matrix in Remark \ref{R:M}. Once one knows $p$, the existence of $\varphi^\perp$
follows from $iii)$ in Lemma \ref{L:nondegen},
and surjectivity is proved.

We are in position to apply the implicit function theorem for any fixed ${R}\in SO(3)$. Actually, by
a compactness argument, we have that there exist $\eps'>0$ and uniquely determined differentiable functions 
$$
\begin{array}{ll}
u:(-\eps',\eps')\times SO(3)\to C^2_{\Sf^2}\!\!\setminus\!\Sf^2~,\quad  &u:(\eps,{R})\mapsto u^\eps_{R}\\
\zeta:(-\eps',\eps')\times SO(3)\to \R^3~,\quad &\zeta:(\eps,{R})\mapsto \zeta^\eps({R})=
(\zeta^\eps_1({R}),\zeta^\eps_2({R}), \zeta^\eps_3({R}))
\end{array}
$$
such that 
$$
\mathcal F(\eps,{R},u^\eps_{R};\zeta^\eps({R}))=0~,
\quad
u^0_{R}=\omega_{\!R}~,\quad \quad\zeta^0({R})=0.
$$
Clearly the function $(\eps,{R})\mapsto u^\eps_{R}$ is differentiable.
Since $\omega_{\!R}$ is embedded, then $u^\eps_{R}$ is embedded as well,
provided that $\eps'$ is small enough. 

Condition $i)$ in the Lemma is fulfilled; $ii)$ follows from 
$\mathcal F_2(\eps,{R},u^\eps_{R};\zeta^\eps({R}))=0$ while
$\mathcal F_1(\eps,{R},u^\eps_{R};\zeta^\eps({R}))=0$ gives $iii)$. 

\medskip

Now we prove that $iv)$ holds for any $\beps\in(0,\eps')$, provided that $\eps'$ is small enough. 
Since $|\omega+e_3|\ge 1$ and $u^\eps_R\to \omega_{\!R}$ uniformly on $\Sf^1$ as $\eps\to 0$, we can assume
that
$$
|u^\eps_R(x)+Re_3|\ge \frac12\quad\text{for any  $x\in\Sf^1, (\eps,R)\in
(-\eps',\eps')\times SO(3)$.}
$$
In particular, Lemma \ref{L:E} guarantees that the function
$\mathcal E^\eps({R})=E_{\eps K}(-Re_3;u^\eps_R)$
is well defined and of class $C^1$ on $SO(3)$, for any $\eps\in (-\eps',\eps')$.
By $iii)$
in Lemma \ref{L:E} we have that the derivative of
$p\mapsto E_{\eps K}(p;u^\eps_R)$ vanishes for $p\in \Sf^2\setminus u^\eps_R(\Sf^1)$,
and we can compute
\begin{equation}
\label{eq:aprile0}
d_R {\mathcal E}^\eps({R})(RT_h)=E'_{\eps K}(-Re_3;u^\eps_R)(d_Ru^\eps_R(RT_h))
\quad\text{for $h\in\{1,2,3\}$},
\end{equation}
where $E'_{\eps K}(-Re_3;\ms\cdot\ms)$ is the differential of the energy
with respect to curves running in $C^2_{\Sf^2\setminus\{-Re_3\}}$.
The $C^1$ dependence of $\mathcal E^\eps({R})$ on $\eps$ and thus
on the pair $(\eps, R)$
is evident. 

\medskip

Next, notice that
$R^\xi_3\omega=\omega\circ\xi$ for any rotation $\xi\in\Sf^1$
(recall that $R^\xi_3$ rotates $\Sf^2$ about the $\langle e_3\rangle$ axis).
Hence
$RR^\xi_3\omega=\omega_{\!R}\circ\xi$ and
$T_{RR^\xi_3\omega}\mathcal S=\big\{\tau\circ\xi~|~\tau\in T_{\omega_{\!R}}\mathcal S~\big\}$
for any $R\in SO(3)$. Taking also $ii), iii)$ into account, we have that 
$$
u^\eps_R\circ\xi\in(T_{RR^\xi_3\omega}\mathcal S)^\perp~,\quad J_\eps(u^\eps_R\circ\xi)=
J_\eps(u^\eps_R)\circ\xi \in T_{RR^\xi_3\omega}\mathcal S\ms.
$$
Since in addition $u^\eps_R\circ\xi$ is close to $\omega_{\!R}\circ\xi=RR^\xi_3\omega$
in the $C^2$-norm by $i)$, we see that    
\begin{equation}
\label{eq:uniqueness}
u^\eps_{RR^\xi_3}=u^\eps_{R}\circ \xi
\end{equation}
by the uniqueness of the function $\eps\mapsto u^\eps_R$ given by the implicit function
theorem. By differentiating (\ref{eq:uniqueness}) with respect to $\xi$ at  $\xi=1$ we obtain
$d_Ru^\eps_R(RT_3)=(u^\eps_R)'$, 
that compared with  (\ref{eq:EK_invariance}) gives
$E'_{\eps K}(-Re_3;u^\eps_R)(d_Ru^\eps_R(RT_3))=E'_{\eps K}(-Re_3;u^\eps_R)(u^\eps_R)'=0$.
Thus $d_R\mathcal E^\eps({R})(RT_3)=0$ by (\ref{eq:aprile0}), and $iv)$ is proved.

\medskip

To prove that $v)$ holds for $\beps$  small enough, first take $R\in SO(3)$, $h\in\{1,2,3\}$
and notice that the condition $u^\eps_R\in T_{\omega_{\!R}}\mathcal S^\perp$ trivially gives
$$
d_R\Big(\fintm u^\eps_R\cdot\ms R(e_j\wedge \omega)\ms dx\Big)(RT_h)=0\ms.
$$
We compute
$d_R R(e_j\wedge \omega)(RT_h)=Re_h\wedge(R(e_j\wedge\omega))
=R\big(e_h\wedge(e_j\wedge\omega)\big)$. Since in addition
$u^\eps_R\cdot R(e_h\wedge(e_j\wedge\omega))=
-(Re_h\wedge u^\eps_R)\cdot (Re_j\wedge \omega_{\!R})
$ we obtain
\begin{equation}
\label{eq:mepsR}
m^\eps_{hj}({R}):=\fintm d_Ru^\eps_R(RT_h)\cdot Re_j\wedge \omega_{\!R}\ms dx=
\fintm (Re_h\wedge u^\eps_R)\cdot (Re_j\wedge\omega_{\!R})\ms dx.
\end{equation}
Since $u^\eps_R\to \omega_{\!R}$ uniformly for $R\in \SO$, from (\ref{eq:mepsR}) we obtain
$$
m^\eps_{hj}({R})=\fintm (Re_h\wedge \omega_{\!R})\cdot (Re_j\wedge\omega_{\!R})\ms dx+o(1)=
m_{hj}+o(1),
$$
where $m_{hj}$ are the entries of the invertible matrix $M$ in Remark \ref{R:M}. 
It follows that the $3\times 3$  matrix
$M^\eps_R=(m^\eps_{hj}({R}))_{j,h=1,2,3}
$
is invertible
for any $R\in\SO$, if $\eps$ is small enough.

We are in position to conclude the proof of $v)$. 
We know that there exists a differentiable function
$(\eps,R)\mapsto \zeta^\eps({R})\in\R^3$
such that 
\begin{equation}
\label{eq:manca}
J_\eps(u^\eps_{R})=\sum_{j=1}^3 \ms \zeta^\eps_j({R})\ms (Re_j\wedge \omega_{\!R}).
\end{equation}
On the other hand, (\ref{eq:aprile0}) and (\ref{eq:EJeps}) give
\begin{equation}
\label{eq:5mesi}
L(u^\eps_R)d_R {\mathcal E}^\eps({R})(RT_h)=
\fintm J_\eps(u^\eps_R)\cdot d_Ru^\eps_R(RT_h)\ms dx,
\end{equation}
by (\ref{eq:manca}) and recalling (\ref{eq:mepsR}) we obtain
$$
L(u^\eps_R)d_R {\mathcal E}^\eps({R})(RT_h)=
\sum_{j=1}^3m^\eps_{hj}({R})\zeta^\eps_j({R})
=\ms e_h\cdot M^\eps_R(\zeta^\eps({R}))\ms.
$$
If $\eps\approx 0$ so that the matrix $M^\eps_R$ is invertible, then 
$R$ is a critical matrix for ${\mathcal E}^\eps$ if and only if $\zeta^\eps({R})=0$, which is equivalent 
to say that $J_\eps(u^\eps_R)=0$.

\medskip
To prove the last claim of the lemma we take $R\in SO(3)$ and compute the Taylor expansion formula of the function
$$
\begin{aligned}
f_R(\eps)=\mathcal E^\eps({R})- \mathcal E^\eps_0({R})=L(u^\eps_R)-1+
\eps\big(A_{K}(-Re_3;u^\eps_R)- A_{K}(-Re_3;\omega_{\!R})\big)
\end{aligned}
$$
at $\eps=0$. Clearly $f_R(0)=0$.  Now we recall that 
$L'(\omega_{\!R})=0$ because $\omega_{\!R}$ is a geodesic, and we write
\begin{multline*}
f'_R(\eps)= \big(L'(u^\eps_R)-L'(\omega_{\!R})\big)(\partial_\eps u^\eps_R)+
\eps\ms A'_{K}(-Re_3;u^\eps_R)(\partial_\eps u^\eps_R)\\
+
\big(A_{K}(-Re_3;u^\eps_R)- A_{K}(-Re_3;\omega_{\!R})\big).
\end{multline*}
To take the limit as $\eps\to0$, we notice that 
$\partial_\eps u^\eps_R$ is uniformly bounded in $C^2_{\Sf^2}$ because
the function $(\eps,R) \mapsto u^\eps_R$ is of class $C^1$. Further,
$L'(u^\eps_R)\to L'(\omega_{\!R})$ in the norm operator,
$A'_K(-Re_3;u^\eps_R)(\partial_\eps u^\eps_R)$ remains bounded
and $A_{K}(-Re_3;u^\eps_R)\to  A_{K}(-Re_3;\omega_{\!R})$.
In conclusion, we have that $f'_R(0)=0$, and therefore $f_R(\eps)=o(\eps)$ as $\eps\to0$,
uniformly on $SO(3)$. That is, (\ref{eq:zero_order}) holds true  ''at the zero order''.

\medskip

To conclude the proof we have to handle the derivatives  of 
$\mathcal E^\eps({R})- \mathcal E^\eps_0({R})$ with respect to $R$, along
any direction $RT_h\in T_{R}SO(3)$. 
We use (\ref{eq:manca}), the second equality in (\ref{eq:mepsR})  and then (\ref{eq:manca}) again  to obtain
\begin{multline*}
\fintm J_\eps(u^\eps_R)\cdot (d_Ru^\eps_R(RT_h))\ms dx=
\sum_{j=1}^3  \zeta^\eps_j({R})\fintm (d_Ru^\eps_R(RT_h))\cdot (Re_j\wedge \omega_{\!R})\ms dx\\
=\sum_{j=1}^3  \zeta^\eps_j({R})\fintm (Re_h\wedge u^\eps_R)\cdot (Re_j\wedge \omega_{\!R})\ms dx=
\fintm J_\eps(u^\eps_R)\cdot (Re_h\wedge u^\eps_R) \ms dx.
\end{multline*}
By (\ref{eq:Jeps0}), the last integral can be written as
$$
\begin{aligned}
\fintm J_0(u^\eps_R)\cdot (Re_h\wedge u^\eps_R)\ms dx
\ms+\ms&
\eps L(u^\eps_R)A'_K(-Re_3;u^\eps_R)(Re_h\wedge u^\eps_R)\\
=&\ms
 \eps \ms L(u^\eps_R) A'_K(-Re_3;u^\eps_R)(Re_h\wedge u^\eps_R)
\end{aligned}
$$
because of (\ref{eq:inv0}).
Thus (\ref{eq:5mesi}) leads to the new formula
$$
d_R {\mathcal E}^\eps({R})(RT_h)
=\eps A'_K(-Re_3;u^\eps_R)(Re_h\wedge u^\eps_R)\ms.
$$
On the other hand, it is easy to see that
$$
d_R {\mathcal E}^\eps_0({R})(RT_h)=\eps  A'_K(-Re_3;\omega_{\!R})(d_R(\omega_{\!R})(RT_h))=
\eps  A'_K(-Re_3;\omega_{\!R})(Re_h\wedge \omega_{\!R}),
$$
because $A_K(\ms\cdot\ms;\omega_{\!R})$ is locally constant, and we can conclude that 
\begin{multline*}
d_R\big( {\mathcal E}^\eps({R})-{\mathcal E}^\eps_0({R})\big)(RT_h)\\ =
\eps\big(A'_K(-Re_3;u^\eps_R)(Re_h\wedge u^\eps_R)-
A'_K(-Re_3;u^\eps_R)(Re_h\wedge \omega_{\!R})\big)=o(\eps),
\end{multline*}
because $u^\eps_R\to \omega_{\!R}$.
The lemma is completely proved.
\QED

\section{Two solutions}
\label{S:two}

In the present section we use Lemma \ref{L:reduction} together with the local variational approach in Section \ref{S:variational}
to provide a more direct, self-contained and analytical treatmen to Viterbo's and Bottkoll's result which avoids the deep and general theories of characteristics and symplectic actions.

We stress the fact that, differently from \cite{G2}, \cite{S0} and \cite{RS}, in the next theorem we do not make any sign assumptions on $K$.
For instance, $K$ might vanish on some geodesic circle of radius 
${{\pi}/{2}}$ about a point $z\in\Sf^2$ and thus
 $\partial{\mathcal D}_{\!\frac{\pi}{2}\!}(z)$
can be  parameterized by two $K$-magnetic geodesics that coincide up to orientation\footnote{\footnotesize{Recall that changing the orientation of a curve only
changes the sign of its curvature.}}. This is the reason why, in that case, we have to add an extra assumption to obtain 
two distinct solutions.
 
\begin{Theorem}
\label{T:main}
Let $K\in C^1(\Sf^2)$ be given. For every $c>0$ large enough, Problem 
{\rm (\ref{P_Arnold})} has at least a solution $\gamma$. 
If in addition $K$ does not vanish on any closed geodesic, or 
\begin{equation}
\label{eq:distinct}
\int\limits_{{\mathcal D}_{\!\frac{\pi}{2}\!}(z)} K(q)\ms d\sigma_{\!q}=\int\limits_{{\mathcal D}_{\!\frac{\pi}{2}\!}(-z)} K(q)\ms d\sigma_{\!q}
\quad
\text{whenever $K\equiv 0$ on ~$\partial{\mathcal D}_{\!\frac{\pi}{2}\!}(z)$,}
\end{equation}
then  for every $c>0$ large enough,  Problem 
(\ref{P_Arnold}) has at least two embedded, distinct solutions.
\end{Theorem}

\proof
Let $\beps$ be given by Lemma \ref{L:reduction}. For any $c > \beps^{-1}$, let $\eps :=  c^{-1} < \beps$ and $(\eps,{R})\mapsto u^\eps_R$, $(\eps,{R})\mapsto \mathcal E^\eps({R})$ be the functions in Lemma \ref{L:reduction}. To every critical point $R^\eps$ for $\mathcal E^\eps$
corresponds a curve $u^\eps_{R^\eps}$ that solves $J_\eps(u^\eps_{R^\eps})=0$. Hence $u^\eps_{R^\eps}$
solves (\ref{eq:kepseq}) and, as explained at the beginning of Section \ref{S:variational}, yields a solution to ($\mathcal P_{K,\eps^{-1}}$) = ($\mathcal P_{K,c}$).

Now, if $\mathcal E^\eps$ is constant, then $u^\eps_{R}$ solves (\ref{eq:kepseq})  for every
$R\in SO(3)$ and the conclusions in Theorem \ref{T:main} hold. Otherwise, take $\underline R^\eps, \overline R^\eps\in SO(3)$ achieving
the minimum and the maximum value of $\mathcal E^\eps$, respectively. Then
$\underline{u}^\eps:=u^\eps_{\underline{R}^\eps}$ and $\overline{u}^\eps:=u^\eps_{\overline{R}^\eps}$ 
solve (\ref{eq:kepseq}) and this concludes the proof of the
existence part.

Next, assume that $\mathcal E^\eps$ is not constant, and that 
$\underline{u}^\eps=\overline{u}^\eps\circ g$ for a diffeomorphism $g$ of $\Sf^1$. 
To conclude the proof we have to show that (\ref{eq:distinct}) can not hold.

We have $E_{\eps K}(\underline{z}^\eps, \underline{u}^\eps)< E_{\eps K}(\overline{z}^\eps, \overline{u}^\eps)$,
that is,
\begin{equation}
\label{eq:dis}
L(\underline{u}^\eps)+\eps A_{K}(\underline{z}^\eps, \underline{u}^\eps)< L(\overline{u}^\eps)+\eps A_{K}(\overline{z}^\eps, \overline{u}^\eps)
\end{equation}
where $\underline{z}^\eps=-\underline{R}^\eps e_3, \overline{z}^\eps=-\overline{R}^\eps e_3$. 
Since
$|(\underline{u}^\eps)'|, |(\overline{u}^\eps)'|$ are constant, then $|g'|$ is constant as well.
Thus 
$|g'|=1$ and $L(\underline{u}^\eps)=L(\overline{u}^\eps)$.
Therefore, (\ref{eq:dis}) implies
\begin{equation}
\label{eq:dis2}
A_{K}(\underline{z}^\eps, \underline{u}^\eps)\neq A_{ K}(\overline{z}^\eps, \overline{u}^\eps)
\end{equation}
for any $\eps\neq 0$. In particular, $g$ can not be a positive rotation of the circle by the property $A2)$ of the area functional. 
Thus $g$ is a counterclockwise rotation of $\Sf^1$. Recall that 
$\underline{u}^\eps$ has curvature $\eps K(\underline{u}^\eps)$ and 
$\overline{u}^\eps$ has curvature $\eps K(\overline{u}^\eps)$. Since changing the orientation of a  curve  
changes the sign of its curvature, we have that at any point $p\in \Gamma:=\underline u^\eps(\Sf^1)=\overline u^\eps(\Sf^1)$
we have $K(p)=-K(p)$. It follows that 
$K\equiv 0$ on $\Gamma$, and hence $\Gamma$
is the boundary of a half-sphere ${\mathcal D}_{\!\frac{\pi}{2}\!}(w^\eps)$. 
We can assume that 
$\underline{u}^\eps$ is a positive parameterization of $\partial {\mathcal D}_{\!\frac{\pi}{2}\!}(w^\eps)$.
Then $\underline{z}^\eps\notin \overline{{\mathcal D}_{\!\frac{\pi}{2}\!}(w^\eps)}$ because 
$\underline{u}^\eps\approx \omega_{\underline{R}^\eps}$, see $i)$ in Lemma \ref{L:reduction}. Next,
since $\overline{u}^\eps$ parameterizes the same geodesic 
with opposite direction, then $\overline{u}^\eps$ 
a positive parameterization of $\partial {\mathcal D}_{\!\frac{\pi}{2}\!}(-w^\eps)$
and $\overline{z}^\eps\notin \overline{{\mathcal D}_{\!\frac{\pi}{2}\!}(-w^\eps)}$.
In particular, from the properties $A3)$ and $A4)$ of the area functional we infer
$$
\begin{aligned}
A_{K}(\underline{z}^\eps, \underline{u}^\eps)=&\ms A_{K}(-w^\eps, \underline{u}^\eps)\ms=\ms
-\frac{1}{2\pi}\int\limits_{{\mathcal D}_{\!\frac{\pi}{2}\!}(w^\eps)}K(q)\ms d\sigma_{\!q}\\
A_{K}(\overline{z}^\eps, \overline{u}^\eps)=&A_{K}(w^\eps, \overline{u}^\eps)=
-\frac{1}{2\pi}\int\limits_{{\mathcal D}_{\!\frac{\pi}{2}\!}(-w^\eps)}K(q)\ms d\sigma_{\!q}\ms,
\end{aligned}
$$
that compared with (\ref{eq:dis2}) shows that (\ref{eq:distinct}) is violated.
The theorem is completely proved.
\QED

\section{Many solutions}
\label{S:many}

In this section we  suggest a way to obtain more and more distinct $K$-magnetic geodesics. It involves the $C^1$
Mel'nikov-type function
\begin{equation}
\label{eq:mel1}
F_{\!K}(z)=\int\limits_{\mathcal D_{\!\frac{\pi}{2}\!}(z)} K({p})\ms d\sigma_{\!p}~,\quad F_{\!K}:\Sf^2\to\R\ms~\!,
\end{equation}
where $K\in C^1(\Sf^2)$ is given. We start by recalling 
the definition of stable critical point proposed in 
\cite[Chapter 2]{AM}, see also \cite{MZh}.

\begin{Definition}
\label{D:def}
Let $\Omega\subset \Sf^2$ be open. We say that ${{F_K}}$ has a stable critical point in $\Omega$ if
there exists $r>0$ such that any  function ${G}\in C^1(\overline \Omega)$
satisfying 
$\displaystyle{\|{G}-{{F_K}}\|_{C^1(\overline \Omega)}<r}$  has a critical point in $\Omega$.
\end{Definition}

If $F_K$ is not constant, then it has at least two distinct stable critical points, namely, its minimum and its maximum.
Different sufficient conditions to have the existence of (possible multiple) stable critical points $z\in \Omega$ for ${F_K}$
are easily given via elementary calculus. 
For instance, one can assume that one of the following conditions holds:
\begin{itemize}
\item[$i)$] $\nabla {{F_K}}(z)\neq 0$ for any $z\in\partial \Omega$, and $\deg(\nabla {{F_K}},\Omega, 0)\neq 0$, 
where ''$\deg$'' is Browder's topological degree;
\item[$ii)$] $\displaystyle{{\min_{\partial \Omega} {{F_K}}>\min_{\Omega} {{F_K}}}}$ ~or~
$\displaystyle{\max_{\partial \Omega} {{F_K}}<\max_{\Omega} {{F_K}}}$;
\item[$iii)$] ${{F_K}}$ is of class $C^2$ on $\Omega$, it has a critical point $z_0\in \Omega$, 
and the Hessian matrix of ${{F_K}}$ at $z_0$ is invertible.
\end{itemize}

In the next result  we show that  any {stable critical point} $z_0$ for $F_K$
gives rise, for any $c>0$ large enough,  to a solution $\gamma^c$ to Problem (\ref{P_Arnold}) which  is a perturbation of the closed geodesic about $z_0$. 
Taking advantage of the remarks at the beginning of Section \ref{S:variational},
we only need to show that for any {stable critical point} $z_0$ for $F_K$
and for any $\eps= c^{-1}\approx 0^+$, there exists a solution $u^\eps$ to (\ref{eq:kepseq}),
such that $u^\eps$ is close to the closed geodesic 
about $z_0$.

\begin{Theorem}
\label{T:main1}
Let $K\in C^1(\Sf^2)$ be given. Assume  that $F_{\!K}$ has a stable 
critical point in an open set $\Omega\subset \Sf^2$, such that $\overline \Omega\subsetneq \Sf^2$. 

Then for every $\eps\in\R$ close enough to $0$, there exists a point $z_\eps\in \Omega$,
an embedding $\omega^\eps:\Sf^1\to\Sf^2$ parameterizing the boundary of a
circle of geodesic radius $\pi/2$ about $z_\eps$, and a solution $u^\eps$ to
($\mathcal P_{K,\eps^{-1}}$), such that
$\|u^\eps-\omega^\eps\|_{C^2}=O(\eps)$. 
\end{Theorem}

\proof
We can assume  $-e_3\notin \overline\Omega$. Otherwise, take any rotation $R\in SO(3)$
such that $-e_3\notin R\overline\Omega$, and look for a solution $\tilde u^\eps$ to
$$
u''+|u'|^2u=L(u)\ms \eps\ms  (K\circ \prescript{t}{}{\!R})(u)~\!u\wedge u'~\qquad \text{on $\Sf^1$},
$$
in a $C^2$-neighborhood of a geodesic circle about some point $\tilde z^\eps\in R\Omega$. Conclude by noticing that
$u^\eps:=\prescript{t}{}{\!R}\tilde u^\eps$ solves (\ref{eq:kepseq}) and approaches
the geodesic circle about $R^{\text{t}} \tilde z^\eps\in \Omega$.

Next, for $z\in\Sf^2\setminus\{-e_3\}$ consider the rotation
$$
N(z)=\left(\begin{array}{ccc}
1-\frac{z_1^2}{1+z_3}& -\frac{z_1z_2}{1+z_3}&z_1\\
&\\
-\frac{z_1z_2}{1+z_3}&1-\frac{z_2^2}{1+z_3}&z_2\\
&\\
-z_1&-z_2&z_3
\end{array}\right)\ms,
$$
that maps $e_3$ to $z$. Clearly the function $N:\Sf^2\setminus\{-e_3\}\to SO(3)$  is
differentiable; its differential $dN(z)$ at any  $z\in\Sf^2\setminus\{-e_3\}$ is a linear map
$T_z\Sf^2\to T_{N(z)}SO(3)$. We have 
\begin{gather}
\label{eq:tan1}
T_z\Sf^2=\langle N(z)e_1,N(z)e_2\rangle\\
\label{eq:tan2}
T_{N(z)}SO(3)=\langle dN(z)\big( N(z)e_1\big),dN(z)\big( N(z)e_2\big)\rangle\oplus\langle N(z)T_3\rangle\ms.
\end{gather}
Equality (\ref{eq:tan1}) and the inclusion $\supseteq$ in (\ref{eq:tan2}) are trivial.
To conclude the proof of (\ref{eq:tan2}) we  need to show that the  matrices 
$$
dN(z)\big( N(z)e_1\big)~,\quad dN(z)\big( N(z)e_2\big)~,\quad N(z)T_3
$$
are linearly independent. By
differentiating
the identity $N(z)e_3=z$ one gets
$$
dN(z)\tau\cdot e_3=\tau~,\quad \tau\in T_z\Sf^2\ms.
$$
By choosing $\tau =N(z)e_h$, $h=1,2$ we infer that the third columns of the matrices 
$dN(z)\big( N(z)e_1\big), dN(z)\big( N(z)e_2\big)$ are linearly independent. Thus the 
matrices $dN(z)\big( N(z)e_1\big)~, dN(z)\big( N(z)e_2\big)$ are linearly independent as well.
On the other hand, the third column on $N(z)T_3$ is identically zero, that concludes the proof of 
(\ref{eq:tan2}).

Now, take the differentiable functions $(\eps,R)\mapsto u^\eps_R\in C^2_{\Sf^2}$, 
$(\eps,R)\mapsto \mathcal E^\eps({R})\in\R$ given by 
Lemma \ref{L:reduction}. To simplify notations, for $z\in \Sf^2\setminus \{-e_3\}$
we write 
$$
\widetilde{\mathcal E^\eps}(z)=\mathcal E^\eps(N(z))=E_{\eps K}(-z;u^\eps_{N(z)})~,\quad
\widetilde{\mathcal E^\eps_0}(z)=\mathcal E^\eps_0(N(z))=E_{\eps K}(-z;N(z)\omega)\ms.
$$
Notice that $N(z)\omega$ parameterizes $\partial {\mathcal D}_{\!\pi/2}(z)$. Therefore,
using  $ii)$ in Lemma \ref{L:E}, property $A4)$ and elementary computations we get
\begin{equation}
\label{eq:const}
\begin{aligned}
\widetilde{\mathcal E^\eps_0}(z)=& L(N(z)\omega)+\eps A_{K}(-z;N(z)\omega) \\
=&
L(\omega)\!-\frac{\eps}{2\pi} \!\!\!\int\limits_{D_{\!\pi/2}(z)}\!\!\!K(q)\ms\! d\sigma_{\!q}=
L(\omega)\!-\frac{\eps}{2\pi} F_{\!K}(z).
\end{aligned}
\end{equation} 
Next, for any small $\eps\neq 0$ consider the function
$$
G^\eps(z)= \frac{2\pi}{\eps}(\widetilde{\mathcal E^\eps}(z)-L(\omega))
$$
and use (\ref{eq:const}) together with $iv)$ in Lemma \ref{L:reduction} to get
$$
\|G^\eps+F_{\!K}\|_{C^1(\overline \Omega)}=
\frac{2\pi}{|\eps|}\big\|E_{\eps K}(-z;u^\eps_{N(z)})-E_{\eps K}(-z;N(z)\omega)\big\|_{C^1(\overline \Omega)}=o(1)
$$
as $\eps\to 0$. 
We see that for $\eps$ small enough the function $G^\eps$
has a critical point $z^\eps\in \Omega$. Thus, for any $\tau\in T_{z^\eps} \Sf^2$ we have
$$
0=d_z\widetilde{\mathcal E^\eps}(z^\eps)\tau=d_R\mathcal E^\eps(N(z^\eps))\big(d_zN(z^\eps)\tau\big)\ms.
$$
Taking (\ref{eq:tan2}) and $iv)$ in Lemma \ref{L:reduction} into account, we infer that 
the matrix $N(z^\eps)$ is critical for ${\mathcal E^\eps}$. Thus,  by arguing as for Theorem \ref{T:main} we 
have that the curve   $u^\eps:=u^\eps_{N(z_\eps)}$ is a solution to ($\mathcal P_{K, \eps^{-1}}$).
\QED

\end{document}